\documentclass[twocolumn,amsmath,aps,fleqn]{revtex4}
\usepackage{graphicx,amssymb}
\begin{document}
\newcommand{\be}{\begin{equation}}
\newcommand{\ee}{\end{equation}}
\newcommand{\bq}{\begin{eqnarray}}
\newcommand{\eq}{\end{eqnarray}}
\newcommand{\bsq}{\begin{subequations}}
\newcommand{\esq}{\end{subequations}}
\newcommand{\bc}{\begin{center}}
\newcommand{\ec}{\end{center}}
\newcommand {\R}{{\mathcal R}}
\newcommand{\al}{\alpha}
\newcommand\lsim{\mathrel{\rlap{\lower4pt\hbox{\hskip1pt$\sim$}}
    \raise1pt\hbox{$<$}}}
\newcommand\gsim{\mathrel{\rlap{\lower4pt\hbox{\hskip1pt$\sim$}}
    \raise1pt\hbox{$>$}}}

\title{Eddington-inspired Born-Infeld gravity: astrophysical and cosmological constraints}

\author{P.P. Avelino}
\email[Electronic address: ]{ppavelin@fc.up.pt}
\affiliation{Centro de Astrof\'{\i}sica da Universidade do Porto, Rua das Estrelas, 4150-762 Porto, Portugal}
\affiliation{Departamento de F\'{\i}sica e Astronomia, Faculdade de Ci\^encias, Universidade do Porto, Rua do Campo Alegre 687, 4169-007 Porto, Portugal}

\date{\today}
\begin{abstract}

In this letter we compute stringent astrophysical and cosmological constraints on a recently proposed Eddington-inspired Born-Infeld theory of gravity. We find, using a generalized version of the Zel'dovich approximation, that in this theory a  pressureless cold dark matter fluid has a non-zero effective sound speed. We compute the corresponding effective Jeans length and show that it is approximately equal to the fundamental length of the theory $R_*=\kappa^{1/2} G^{-1/2}$, where $\kappa$ is the only additional parameter of theory with respect to general relativity and $G$ is the gravitational constant. This scale determines the minimum size of compact objects which are held together by gravity. We also estimate the critical mass above which pressureless compact objects are unstable to colapse into a black hole, showing that it is approximately equal to the fundamental mass $M_* = \kappa^{1/2} c^2 G^{-3/2}$, and we show that the maximum density attainable inside stable compact stars is roughly equal to the fundamental density $\rho_*=\kappa^{-1} c^2$, where $c$ is the speed of light in vacuum. We find that the mere existence of astrophysical objects of size $R$ which are held together by their own gravity leads to the constraint $\kappa < G R^2$. In the case of neutron stars this implies that $\kappa < 10^{-2} \, {\rm m^5 \, kg^{-1} \, s^{-2}}$, a limit which is stronger by about $10$ orders of magnitude than big bang nucleosynthesis constraints and by more than $7$ orders of magnitude than solar constraints.

\end{abstract}
\maketitle

\section{\label{intr}Introduction}

Recently, a new Eddington-inspired theory of gravity with a Born-Infeld like structure \cite{Born:1934gh} has been proposed by Banados and Ferreira \cite{Banados:2010ix} (see also \cite{Deser:1998rj,Vollick:2003qp,Wohlfarth:2003ss,Nieto:2004qj,Comelli:2005tn,Vollick:2005gc,Zinoviev:2006,Ferraro:2008ey,Fiorini:2009ux,Ferraro:2009zk,Gullu:2010pc,Alishahiha:2010iq,Gullu:2010em} for other relevant studies of Born-Infeld type gravitational models and \cite{Clifton:2011jh} for a recent review on alternative theories of gravity). A key feature of the Eddington-inspired Born-Infeld (EiBI) theory of gravity introduced in \cite{Banados:2010ix} is that it is equivalent to Einstein's general relativity in vacuum. In \cite{Banados:2010ix} it was shown that, in the non-relativistic limit, the EiBI theory of gravity leads to a modified Poisson equation given by
\be
\nabla^2 \phi = 4 \pi G \rho_m+\frac{\kappa}{4}\nabla^2 \rho_m\,,
\label{poisson}
\ee
where $\phi$ is the gravitational potential and $\rho_m$ is the matter density. The gravitational constant $G$, $\kappa$ and the speed of light in vacuum $c$ are the only parameters of the theory. With these parameters it is possible to define a fundamental length, time, mass and density respectively by
\be
R_*=\sqrt{\frac{\kappa}{G}}\,,\ \ t_*=\sqrt{\frac{\kappa}{G c^2}}\,,\ \ M_*=\sqrt{\frac{\kappa c^4}{G^3}}\,,\ \ \rho_*=\frac{c^2}{\kappa}\,,
\label{fstar}
\ee
whose physical interpretation will be revealed in this letter. In \cite{Banados:2010ix} it was shown that the EiBI theory of gravity significantly changes the universe dynamics at early times, leading to a non-singular cosmological model. In \cite{Pani:2011mg} the EiBI theory has also been shown to support compact stars made of pressureless dust. Astrophysical constraints on the single extra parameter of the theory $\kappa$ have been determined considering the physics of astronomical objects such as neutron stars \cite{Pani:2011mg} and the Sun \cite{Casanellas:2011kf}.

In this letter we compute both astrophysical and cosmological constraints on the EiBI theory of gravity. In Sec. II basic results for the evolution of the Universe during the radiation era are reviewed and a constraint on the value of $\kappa$ is derived from primordial nucleosynthesis. In Sec. III the Zel'dovich approximation is generalized to account for the modification to the Poisson equation in Eq. (\ref{poisson}) which it is shown to lead to an effective fundamental effective length, in the case of pressureless cold dark matter.  We use this result to compute the minimum size (allowed by the theory) of compact objects which are held together by gravity and to determine the maximum density attainable inside stable compact stars. We also obtain the critical mass above which pressureless compact stars cannot exist. We then use these results to derive stringent astrophysical constraints on the value of $\kappa$. We conclude in Sec. IV.

\section{Background evolution of the Universe}

The EiBI theory of gravity leads to modifications to the dynamics of the Universe at early times. In \cite{Banados:2010ix} it has been shown that, in the radiation era, the Friedmann equation is given by
\be
H^2=\frac{8\pi G c^2}{3\kappa} \times f({\tilde \rho})\,,
\ee
where
\bq
f({\tilde \rho})&=&\left({\tilde \rho}-1+\frac{(1+{\tilde \rho})^{1/2}(3-{\tilde \rho})^{3/2}}{3^{3/2}}\right)\times \nonumber \\  
&\times& \frac{(1+{\tilde \rho})(3-{\tilde \rho})^2}{(3+{\tilde \rho^2})^2}\,,
\eq
${\tilde \rho}=\kappa \rho /c^2$, $H={\dot a}/a$ is the Hubble parameter, $a$ is the cosmological scale factor and a dot represents a derivative with respect to the physical time $t$. Two stationary points with $H=0$ at ${\tilde \rho}=3$ and ${\tilde \rho}=-1$ were identified, for positive and negative $\kappa$, respectively. For small ${\tilde \rho}$ one recovers the usual dynamics with 
\be
H^2=8\pi G \rho/3\,,
\ee
but for values of ${\tilde \rho}$ of order unity the dynamics is severely modified. In this letter we shall assume that $\kappa > 0$, so that ${\tilde \rho} > 0$, in order to avoid instabilities associated with an imaginary effective sound speed (see Sec. III). 

\subsection{Primordial nucleosynthesis constraint}

Less than one second after the big bang protons and neutrons were in thermal equilibrium. However, about one second after the big bang the temperature $T$ drops below $1 \, {\rm MeV}$ and the neutron-proton inter-conversion rate ($\sim G_F^2 T^5$, where $G_F$ is the Fermi coupling constant) becomes smaller than expansion rate ($H$). This leads to the approximate freeze out of the ratio between the number of neutrons and the number of protons $t_{nuc} \sim 1 \, {\rm s}$ after the big bang, except for free neutron decay. Although the formation of $^4 He$ is delayed until the temperature of the Universe is low enough for deuterium to form (at about $T=0.1 \, {\rm MeV}$), it is crucial that the dynamics of the Universe at the start of the primordial nucleosynthesis epoch ($t_{nuc} \sim 1 \, {\rm s}$) is very close to the standard one in order that the good agreement between primordial nucleosynthesis predictions and observed light element abundances is preserved (see \cite{Nakamura} for a recent review). This implies that the following very conservative constraint must be satisfied
\be
{\tilde \rho}_{nuc}=\frac{\kappa \rho_{nuc}}{c^2} < 3\,,
\label{nuccons1}
\ee
with $ \rho_{nuc}=\rho(t_{nuc})$. Hence, taking into account that $\rho_{nuc} \sim 3H_{nuc}^2/(8\pi G)$, Eq. (\ref{nuccons1}) implies that
\be
\kappa < 8\pi G R_{Hnuc}^2 \sim 6 \times 10^8 \, {\rm m^5 \, kg^{-1} \, s^{-2}}\,,
\label{nuccons2}
\ee
where $H_{nuc}=H(t_{nuc})$, the Hubble radius is defined by $R_H=c H^{-1}$, and its value at the nucleosynthesis epoch is $R_{Hnuc} \sim R_\odot \sim 2 \, {\rm light \ seconds}$ ($R_\odot$ is the solar radius). The constraint in Eq. (\ref{nuccons2}) may be improved by almost an order of magnitude by requiring that the variation of the Hubble parameter at $t_{nuc} \sim 1 \, {\rm s}$, with respect to the standard model value, is less than $10 \, \%$. Primordial nucleosynthesis provides the strongest cosmological constraint on the EiBI theory of gravity. In the next section we shall show that the constraint in Eq. (\ref{nuccons2}) might be improved by several orders of magnitude by taking into account the mere existence of compact round astronomical objects. 

\section{Linear evolution of density perturbations}

In this section we shall consider the non-relativistic regime and generalize the Zel'dovich approximation \cite{Zeld}, for the evolution of cold dark matter density perturbations in an expanding background, to account for the modification to the Poisson equation in Eq. (\ref{poisson}). Here we shall consider times much later than $t_{nuc}$ and consequently the background evolution of the Universe is that of the standard cosmological model. 

The trajectory of the cold dark matter particles in a homogeneous and isotropic Friedmann-Robertson-Walker background can be written as
\be
{\vec r}=a(t)\left[{\vec q}+{\vec \psi}({\vec q},t)\right].
\label{eq:position}
\ee
where ${\vec q}$ is the unperturbed comoving position and ${\vec \psi}({\vec q},t)$ is the comoving displacement. Calculating the first derivative of Eq. (\ref{eq:position}) with respect to the physical time one obtains
\be
{\vec v}=H{\vec r} +{\vec v}_{pec},
\label{eq:velocity}
\ee
where ${\vec v}={\dot {\vec r}}$ and ${\vec v}_{pec}=a {\dot {\vec \psi}}$ is the peculiar velocity. 

The gravitational acceleration felt by the cold dark matter particles is equal to
\be
{\vec {\rm a}} \equiv {\ddot r}= -\nabla \phi.
\label{eq:accm}
\ee
where $\phi$ is given by a generalized Poisson equation
\be
\nabla^2 \phi = 4 \pi G ({\bar \rho}+3{\bar p}+\delta \rho_m)+\frac{\kappa}{4}\nabla^2 \rho_m,
\label{eq:poisson}
\ee
where $\delta \rho_m \equiv \rho_m-{\bar \rho}_m$, $\rho_m$ is the matter density,  ${\bar \rho}_m$ is the average matter density, ${\bar \rho}$ is the average density and ${\bar p}$ is the average pressure. Eq. (\ref{eq:poisson}) generalizes Eq. (\ref{poisson}) to account for a homogeneous cosmological background with an arbitrary equation of state. However, the two equations are equivalent if ${\bar p}=0$. 

Mass conservation requires that
\be
{\bar \rho}_m a^3 d^3 q = \rho_m d^3 r,
\label{eq:mconserv}
\ee
where the infinitesimal volume elements $d^3 r$ and $d^3 q$ are related by
\bq
\frac{d^3 r}{d^3 q}&=&\left|\frac{\partial {\vec r}}{\partial {\vec q}}\right|=a^3\left(1+\sum_i \psi_{i,i} + ...\right) \sim \nonumber\\
&\sim& a^3(1+a \nabla \cdot {\vec \psi}),
\label{eq:jacobian1}
\eq
a comma denotes a partial derivative with respect to the comoving coordinates and the approximation is valid up to first order in the comoving displacement. Hence 
\be
\delta \sim - a \nabla \cdot {\vec \psi}\,,
\label{eq:zel1}
\ee
where  $\delta \equiv \delta \rho_m/{\bar \rho}_m$.
Integrating Eq. (\ref{eq:poisson}) one obtains the first order solution
\be
\nabla \phi = \frac {4 \pi G ({\bar \rho}+3{\bar p})}{3} {\vec r} - 
4\pi G {\bar \rho}_m a {\vec \psi}_\parallel +\frac{\kappa {\bar \rho}_m}{4}\nabla \delta ,
\label{eq:poissonint}
\ee
with ${\vec \psi}= {\vec \psi}_\parallel+ {\vec \psi}_\perp$ where ${\vec \psi}_\parallel$ and ${\vec \psi}_\perp$ are the irrotational and divergence parts of the comoving displacement, respectively ($\nabla \times {\vec \psi}_\parallel= {\vec 0}$ and $\nabla \cdot {\vec \psi}_\perp = 0$). A more general solution may be obtained by adding, to the right hand side of  Eq. (\ref{eq:poisson}), the term $\nabla \varphi$ where $\varphi$ is an arbitrary scalar field satisfying the Laplace equation $\nabla^2 \varphi=0$. For simplicity, we shall ignore that extra term since it will not have any impact on our conclusions. Using the Rachaudhury equation
\be
\frac{\ddot a}{a}=-\frac{4\pi G ({\bar \rho}+3{\bar p})}{3},
\label{eq:rach}
\ee
together with Eqs. (\ref{eq:position}), (\ref{eq:accm}) and (\ref{eq:poissonint}) one may show that
\be
{\ddot {\vec \psi}}+2H {\dot {\vec \psi}} -4\pi G {\bar \rho}_m 
{\vec \psi}_\parallel = -\frac{\kappa {\bar \rho}_m}{4 a}\nabla \delta.
\label{eq:zel2}
\ee
Decomposing Eq. (\ref{eq:zel2}) into its parallel and perpendicular components one finds
\bq
{\ddot {\vec \psi}_\parallel}&+&2H {\dot {\vec \psi}_\parallel} -4\pi G {\bar \rho}_m {\vec \psi}_\parallel = -\frac{1}{a} \frac{\kappa}{4}\nabla \rho_m\,,
\label{eq:zelparallel} \\
{\ddot {\vec \psi}_\perp}&+&2H {\dot {\vec \psi}_\perp} 
= 0\,.
\label{eq:zelperp}
\eq
Integrating Eq. (\ref{eq:zelperp}) one finds that ${\vec v}_{pec \perp}=a {\dot {\vec \psi}}_{pec \perp} \propto a^{-1}$, which simply expresses the conservation of angular momentum. On the other hand, using Eqs. (\ref{eq:zel1}) and (\ref{eq:zel2}) one finally obtains an equation for the evolution of the cold dark matter density perturbations 
\be
{\ddot \delta}+2H {\dot \delta} -4\pi G {\bar \rho}_m \delta =c_{seff}^2 \nabla^2 \delta\,,
\label{eq:denevol}
\ee
where $c_{seff}^2=\kappa {\bar \rho}_m/4$ is the effective sound speed squared of the cold dark matter in the EiBI theory of gravity.  

In Fourier space one obtains 
\be
{\ddot \delta}_{\vec k}+2H {\dot \delta}_{\vec k} 
-\left(4\pi G {\bar \rho}_m -\frac{k^2 c_{seff}^2}{a^2}\right) \delta_{\vec k} = 0.
\label{eq:denevfour}
\ee
where ${\vec k}$ is the comoving wavenumber and $k=|{\vec k}|$. In this letter we assume that $c_{seff}^2 > 0$, or equivalently $\kappa > 0$, in order to avoid unwanted instabilities associated with an imaginary effective sound speed.

\subsection{Astrophysical constraints}

An effective Jeans length can be defined for the cold dark matter as 
\be
\lambda_{Jeff}= \frac{2 \pi a}{k_{Jeff}}=c_{seff} {\sqrt{\frac{ \pi}{G {\bar \rho}_m}}}= {\sqrt{\frac{\pi \kappa}{4G}}} \sim R_*\,,
\label{eq:jeans}
\ee
where $k_{Jeff}$ is the value of $k$ for which the term within brackets in Eq. ({\ref{eq:denevfour}) is equal to zero.
The effective Jeans length for the cold dark matter in the EiBI theory of gravity defines the critical scale bellow which the collapse of pressureless dust is no longer possible (for wavelengths $\lambda < \lambda_{Jeff} \sim R_*$ matter fields oscillate coherently). Consequently, $\lambda_{Jeff}$ determines the minimum scale of compact objects which are held together by gravity (note that including pressure increases the Jeans scale). The demonstration that $\lambda_{Jeff}$ is independent of the matter density being approximately equal to the fundamental scale of the theory $R_*$ is one of the key results of this letter.

By requiring that $\lambda_{Jeff}$ is equal to the schwarzchild radius $r_s$,
\be
\lambda_{Jeff} =  {\sqrt{\frac{\pi \kappa}{4G}}} = \frac{2GM}{c^2} =r_s\,,
\ee
one obtains the critical mass above which pressureless compact stars are unstable to collapse into a black hole 
\be
M = \frac{{\sqrt{\pi}}}{4} \kappa^{1/2} c^2 G^{-3/2} \sim M_*\,,
\ee
which is essentially equal to the fundamental mass of the theory $M_*$. The fundamental density, given by
\be
\rho_*= \frac{M_*}{R_*^3}=\frac{c^2}{\kappa} \sim \frac{c^2}{G} \lambda_{eff}^{-2}\,,
\ee
provides an estimate of the maximum density attainable inside stable compact stars (note that adding pressure increases the Jeans scale, leading to a decrease of the maximum density with respect to the pressureless case).

By requiring that $\lambda_{Jeff}$ is smaller than the solar radius $R_\odot$ one obtains the following conservative constraint 
\be
\kappa < \frac{4}{\pi} G R_\odot^2 \sim 3 \times 10^{7} \, {\rm m^5 \, kg^{-1} \, s^{-2}}\,, \label{astcons}
\ee
which is about two orders of magnitude weaker than that derived in \cite{Casanellas:2011kf}, where a detailed model for the structure of the sun has been considered. However, much stronger constraints on $\kappa$ may be obtained by considering smaller astrophysical objects.

There are several natural satellites in the Solar System which are massive enough to relax to a rounded shape through their internal gravity. Some of them have radii of the order of $100 \, {\rm km}$ \cite{BonnieJ2003329}. Substituting $R_\odot$ by $R =100 \, {\rm km}$ in Eq. (\ref{astcons}) improves the $\kappa$ constraint by more than $7$ orders of magnitude. On the other hand, neutron stars \cite{Lattimer:2004pg} with a typical radius of about $R_{NS} \sim 12 \, {\rm km}$ (nearly 5 orders of magnitude smaller than $R_\odot$) are also held together by gravity. They have core densities which are predicted to be larger than $\rho_c \sim 10^{17} \, {\rm kg \, m^3}$. By requiring that $\rho_* > 10^{17} \, {\rm kg \, m^3}$ one obtains the constraint $\kappa < 1 \, {\rm m^5 \, kg^{-1} \, s^{-2}}$ which is similar to the one obtained in \cite{Pani:2011mg} considering a relativistic model for internal structure of the neutron star. An even tighter constraint may be obtained by requiring that the minimum scale of compact objects which are held together by gravity $\lambda_{Jeff} \sim R_*$ is smaller than $12 \, {\rm km}$. Substituting $R_\odot$ by $R =12 \, {\rm km}$ in Eq. (\ref{astcons}) one obtains the constraint
\be
\kappa < 10^{-2} \, {\rm m^5 \, kg^{-1} \, s^{-2}}\,,\label{tcons}
\ee
which is more than $9$ orders of magnitude stronger than the limit given in Eq. (\ref{astcons}) and more than $7$ orders of magnitude tighter than the solar constraints obtained in \cite{Casanellas:2011kf}. Eq. (\ref{tcons}) also strengthens the constraint given in \cite{Pani:2011mg} by two orders of magnitude.

\section{\label{conc}Conclusions}

In this letter we determined astrophysical and cosmological constraints on a recently proposed EiBI theory of gravity. We have shown, using a generalized version of the Zel'dovich approximation, that in this theory a pressureless cold dark matter fluid has a non-zero effective Jeans length. We used this result to provide a physical interpretation of the fundamental length $R_*$, mass $M_*$ and density $\rho_*$ of the theory and to obtain stringent limits on $\kappa$, the only additional parameter of theory with respect to general relativity. 

The cosmological and astrophysical constraints can be summarized as $\kappa \lsim G R^2$, where $R$ is either the Hubble radius at the onset of primordial nucleosynthesis ($t_{nuc} \sim 1 \, {\rm s}$) or the scale of compact objects which are held  together by gravity. The strongest astrophysical limit ($\kappa < 10^{-2} \, {\rm m^5 \, kg^{-1} \, s^{-2}}$) is about $10$ orders of magnitude  stronger than big bang nucleosynthesis constraints, yielding a constraint on the fundamental mass ($M_* < 5 \, M_\odot$) and density ($\rho_* > 9 \times 10^{18} \, {\rm kg \, m^{-3}}$) of the theory. These limits imply that large changes with respect to the dynamics of the standard cosmological model in the early Universe are expected in the context of the EiBI theory of gravity but only for cosmic times $t < 10^{-5} \, {\rm s}$.

\begin{acknowledgments}

We thank Margarida Cunha for useful comments on the manuscript. This work is partially supported by FCT-Portugal through the project CERN/FP/116358/2010.

\end{acknowledgments}


\bibliography{EiBI}

\begin{thebibliography}{22}
\expandafter\ifx\csname natexlab\endcsname\relax\def\natexlab#1{#1}\fi
\expandafter\ifx\csname bibnamefont\endcsname\relax
  \def\bibnamefont#1{#1}\fi
\expandafter\ifx\csname bibfnamefont\endcsname\relax
  \def\bibfnamefont#1{#1}\fi
\expandafter\ifx\csname citenamefont\endcsname\relax
  \def\citenamefont#1{#1}\fi
\expandafter\ifx\csname url\endcsname\relax
  \def\url#1{\texttt{#1}}\fi
\expandafter\ifx\csname urlprefix\endcsname\relax\def\urlprefix{URL }\fi
\providecommand{\bibinfo}[2]{#2}
\providecommand{\eprint}[2][]{\url{#2}}

\bibitem[{\citenamefont{Born and Infeld}(1934)}]{Born:1934gh}
\bibinfo{author}{\bibfnamefont{M.}~\bibnamefont{Born}} \bibnamefont{and}
  \bibinfo{author}{\bibfnamefont{L.}~\bibnamefont{Infeld}},
  \bibinfo{journal}{Proc.Roy.Soc.Lond.} \textbf{\bibinfo{volume}{A144}},
  \bibinfo{pages}{425} (\bibinfo{year}{1934}).

\bibitem[{\citenamefont{Banados and Ferreira}(2010)}]{Banados:2010ix}
\bibinfo{author}{\bibfnamefont{M.}~\bibnamefont{Banados}} \bibnamefont{and}
  \bibinfo{author}{\bibfnamefont{P.~G.} \bibnamefont{Ferreira}},
  \bibinfo{journal}{Phys.Rev.Lett.} \textbf{\bibinfo{volume}{105}},
  \bibinfo{pages}{011101} (\bibinfo{year}{2010}), \eprint{1006.1769}.

\bibitem[{\citenamefont{Deser and Gibbons}(1998)}]{Deser:1998rj}
\bibinfo{author}{\bibfnamefont{S.}~\bibnamefont{Deser}} \bibnamefont{and}
  \bibinfo{author}{\bibfnamefont{G.}~\bibnamefont{Gibbons}},
  \bibinfo{journal}{Class.Quant.Grav.} \textbf{\bibinfo{volume}{15}},
  \bibinfo{pages}{L35} (\bibinfo{year}{1998}), \eprint{hep-th/9803049}.

\bibitem[{\citenamefont{Vollick}(2004)}]{Vollick:2003qp}
\bibinfo{author}{\bibfnamefont{D.~N.} \bibnamefont{Vollick}},
  \bibinfo{journal}{Phys.Rev.} \textbf{\bibinfo{volume}{D69}},
  \bibinfo{pages}{064030} (\bibinfo{year}{2004}), \eprint{gr-qc/0309101}.

\bibitem[{\citenamefont{Wohlfarth}(2004)}]{Wohlfarth:2003ss}
\bibinfo{author}{\bibfnamefont{M.~N.} \bibnamefont{Wohlfarth}},
  \bibinfo{journal}{Class.Quant.Grav.} \textbf{\bibinfo{volume}{21}},
  \bibinfo{pages}{1927} (\bibinfo{year}{2004}), \eprint{hep-th/0310067}.

\bibitem[{\citenamefont{Nieto}(2004)}]{Nieto:2004qj}
\bibinfo{author}{\bibfnamefont{J.}~\bibnamefont{Nieto}},
  \bibinfo{journal}{Phys.Rev.} \textbf{\bibinfo{volume}{D70}},
  \bibinfo{pages}{044042} (\bibinfo{year}{2004}), \eprint{hep-th/0402071}.

\bibitem[{\citenamefont{Comelli}(2005)}]{Comelli:2005tn}
\bibinfo{author}{\bibfnamefont{D.}~\bibnamefont{Comelli}},
  \bibinfo{journal}{Phys.Rev.} \textbf{\bibinfo{volume}{D72}},
  \bibinfo{pages}{064018} (\bibinfo{year}{2005}), \eprint{gr-qc/0505088}.

\bibitem[{\citenamefont{Vollick}(2005)}]{Vollick:2005gc}
\bibinfo{author}{\bibfnamefont{D.~N.} \bibnamefont{Vollick}},
  \bibinfo{journal}{Phys.Rev.} \textbf{\bibinfo{volume}{D72}},
  \bibinfo{pages}{084026} (\bibinfo{year}{2005}), \eprint{gr-qc/0506091}.

\bibitem[{\citenamefont{Zinoviev}(2006)}]{Zinoviev:2006}
\bibinfo{author}{\bibfnamefont{Y.~M.} \bibnamefont{Zinoviev}},
  \bibinfo{journal}{Journal of High Energy Physics}
  \textbf{\bibinfo{volume}{2006}}, \bibinfo{pages}{009} (\bibinfo{year}{2006}).

\bibitem[{\citenamefont{Ferraro and Fiorini}(2008)}]{Ferraro:2008ey}
\bibinfo{author}{\bibfnamefont{R.}~\bibnamefont{Ferraro}} \bibnamefont{and}
  \bibinfo{author}{\bibfnamefont{F.}~\bibnamefont{Fiorini}},
  \bibinfo{journal}{Phys.Rev.} \textbf{\bibinfo{volume}{D78}},
  \bibinfo{pages}{124019} (\bibinfo{year}{2008}), \eprint{0812.1981}.

\bibitem[{\citenamefont{Fiorini and Ferraro}(2009)}]{Fiorini:2009ux}
\bibinfo{author}{\bibfnamefont{F.}~\bibnamefont{Fiorini}} \bibnamefont{and}
  \bibinfo{author}{\bibfnamefont{R.}~\bibnamefont{Ferraro}},
  \bibinfo{journal}{Int.J.Mod.Phys.} \textbf{\bibinfo{volume}{A24}},
  \bibinfo{pages}{1686} (\bibinfo{year}{2009}), \eprint{0904.1767}.

\bibitem[{\citenamefont{Ferraro and Fiorini}(2010)}]{Ferraro:2009zk}
\bibinfo{author}{\bibfnamefont{R.}~\bibnamefont{Ferraro}} \bibnamefont{and}
  \bibinfo{author}{\bibfnamefont{F.}~\bibnamefont{Fiorini}},
  \bibinfo{journal}{Phys.Lett.} \textbf{\bibinfo{volume}{B692}},
  \bibinfo{pages}{206} (\bibinfo{year}{2010}), \eprint{0910.4693}.

\bibitem[{\citenamefont{Gullu et~al.}(2010{\natexlab{a}})\citenamefont{Gullu,
  Sisman, and Tekin}}]{Gullu:2010pc}
\bibinfo{author}{\bibfnamefont{I.}~\bibnamefont{Gullu}},
  \bibinfo{author}{\bibfnamefont{T.~C.} \bibnamefont{Sisman}},
  \bibnamefont{and} \bibinfo{author}{\bibfnamefont{B.}~\bibnamefont{Tekin}},
  \bibinfo{journal}{Class.Quant.Grav.} \textbf{\bibinfo{volume}{27}},
  \bibinfo{pages}{162001} (\bibinfo{year}{2010}{\natexlab{a}}),
  \eprint{1003.3935}.

\bibitem[{\citenamefont{Alishahiha et~al.}(2010)\citenamefont{Alishahiha,
  Naseh, and Soltanpanahi}}]{Alishahiha:2010iq}
\bibinfo{author}{\bibfnamefont{M.}~\bibnamefont{Alishahiha}},
  \bibinfo{author}{\bibfnamefont{A.}~\bibnamefont{Naseh}}, \bibnamefont{and}
  \bibinfo{author}{\bibfnamefont{H.}~\bibnamefont{Soltanpanahi}},
  \bibinfo{journal}{Phys.Rev.} \textbf{\bibinfo{volume}{D82}},
  \bibinfo{pages}{024042} (\bibinfo{year}{2010}), \eprint{1006.1757}.

\bibitem[{\citenamefont{Gullu et~al.}(2010{\natexlab{b}})\citenamefont{Gullu,
  Sisman, and Tekin}}]{Gullu:2010em}
\bibinfo{author}{\bibfnamefont{I.}~\bibnamefont{Gullu}},
  \bibinfo{author}{\bibfnamefont{T.~C.} \bibnamefont{Sisman}},
  \bibnamefont{and} \bibinfo{author}{\bibfnamefont{B.}~\bibnamefont{Tekin}},
  \bibinfo{journal}{Phys.Rev.} \textbf{\bibinfo{volume}{D82}},
  \bibinfo{pages}{124023} (\bibinfo{year}{2010}{\natexlab{b}}),
  \eprint{1010.2411}.

\bibitem[{\citenamefont{Clifton et~al.}(2011)\citenamefont{Clifton, Ferreira,
  Padilla, and Skordis}}]{Clifton:2011jh}
\bibinfo{author}{\bibfnamefont{T.}~\bibnamefont{Clifton}},
  \bibinfo{author}{\bibfnamefont{P.~G.} \bibnamefont{Ferreira}},
  \bibinfo{author}{\bibfnamefont{A.}~\bibnamefont{Padilla}}, \bibnamefont{and}
  \bibinfo{author}{\bibfnamefont{C.}~\bibnamefont{Skordis}}
  (\bibinfo{year}{2011}), \eprint{1106.2476}.

\bibitem[{\citenamefont{Pani et~al.}(2011)\citenamefont{Pani, Cardoso, and
  Delsate}}]{Pani:2011mg}
\bibinfo{author}{\bibfnamefont{P.}~\bibnamefont{Pani}},
  \bibinfo{author}{\bibfnamefont{V.}~\bibnamefont{Cardoso}}, \bibnamefont{and}
  \bibinfo{author}{\bibfnamefont{T.}~\bibnamefont{Delsate}},
  \bibinfo{journal}{Phys.Rev.Lett.} \textbf{\bibinfo{volume}{107}},
  \bibinfo{pages}{031101} (\bibinfo{year}{2011}), \eprint{1106.3569}.

\bibitem[{\citenamefont{Casanellas et~al.}(2012)\citenamefont{Casanellas, Pani,
  Lopes, and Cardoso}}]{Casanellas:2011kf}
\bibinfo{author}{\bibfnamefont{J.}~\bibnamefont{Casanellas}},
  \bibinfo{author}{\bibfnamefont{P.}~\bibnamefont{Pani}},
  \bibinfo{author}{\bibfnamefont{I.}~\bibnamefont{Lopes}}, \bibnamefont{and}
  \bibinfo{author}{\bibfnamefont{V.}~\bibnamefont{Cardoso}},
  \bibinfo{journal}{Astrophys.J.} \textbf{\bibinfo{volume}{745}},
  \bibinfo{pages}{15} (\bibinfo{year}{2012}), \eprint{1109.0249}.

\bibitem[{\citenamefont{Nakamura and Group}(2010)}]{Nakamura}
\bibinfo{author}{\bibfnamefont{K.}~\bibnamefont{Nakamura}} \bibnamefont{and}
  \bibinfo{author}{\bibfnamefont{P.~D.} \bibnamefont{Group}},
  \bibinfo{journal}{Journal of Physics G: Nuclear and Particle Physics}
  \textbf{\bibinfo{volume}{37}}, \bibinfo{pages}{075021}
  (\bibinfo{year}{2010}).

\bibitem[{\citenamefont{{Zel'Dovich}}(1970)}]{Zeld}
\bibinfo{author}{\bibfnamefont{Y.~B.} \bibnamefont{{Zel'Dovich}}},
  \bibinfo{journal}{Astron. \& Astrophys.} \textbf{\bibinfo{volume}{5}},
  \bibinfo{pages}{84} (\bibinfo{year}{1970}).

\bibitem[{\citenamefont{J. and Buratti}(2003)}]{BonnieJ2003329}
\bibinfo{author}{\bibfnamefont{B.}~\bibnamefont{J.}} \bibnamefont{and}
  \bibinfo{author}{\bibnamefont{Buratti}}, in
  \emph{\bibinfo{booktitle}{Encyclopedia of Physical Science and Technology
  (Third Edition)}}, edited by
  \bibinfo{editor}{\bibfnamefont{E.}~\bibnamefont{in~Chief: Robert A.~Meyers}}
  (\bibinfo{publisher}{Academic Press}, \bibinfo{address}{New York},
  \bibinfo{year}{2003}), pp. \bibinfo{pages}{329 -- 355},
  \bibinfo{edition}{third edition} ed.

\bibitem[{\citenamefont{Lattimer and Prakash}(2004)}]{Lattimer:2004pg}
\bibinfo{author}{\bibfnamefont{J.}~\bibnamefont{Lattimer}} \bibnamefont{and}
  \bibinfo{author}{\bibfnamefont{M.}~\bibnamefont{Prakash}},
  \bibinfo{journal}{Science} \textbf{\bibinfo{volume}{304}},
  \bibinfo{pages}{536} (\bibinfo{year}{2004}), \eprint{astro-ph/0405262}.

\end{thebibliography}

\end{document}